\documentclass[12pt]{article}
\usepackage{amssymb,latexsym,amsmath,amsfonts,multicol}
\usepackage[margin=1in]{geometry}
\usepackage{graphicx}
\usepackage{float}
\date{}
\begin{document}
\title{Exact Solutions and Symmetry Analysis of a Boussinesq type Equation for Longitudinal Waves through a Magneto-Electro-Elastic Circular Rod}
\author{Arindam Ghosh\footnote{email: arindamghosh227@gmail.com},  Sarit Maitra\footnote{email: sarit2010.nt@gmail.com}\\
\newline Department of Mathematics\\ NIT Durgapur, India\\  \\ 
      Asesh Roy Chowdhury\footnote{email: arc.roy@gmail.com},\\Department of Physics\\
         Jadavpur University, India }     
\maketitle
\begin{abstract}
In this work the longitudinal wave equation through magneto-electro-elastic circular rod is studied analytically. Painlev\'e test is performed to check integrability of the equation. Exact solitary wave solutions are found by homogeneous balance method and Tanh method and are plotted using suitable values of physical parameters. A similar solution to an earlier related work [17, 20] is obtained. However we have obtained a new  solution as well. Lie symmetry analysis has been done and similarity reduction of the equation is presented. Numerical results related to phase velocity etc for different materials has been derived and compared.
\end{abstract}
\textbf{PACS Nos 02.20.Sv; 02.30.Ik; 02.30.Jr; 05.45.Yv}
\newpage
\section{Introduction:}
~~~~~~~~     Nonlinear partial differential equations(PDE) are rife in contemporary mathematical modeling. Many problems of physics, mathematics, biology, chemistry are described by nonlinear differential equations [1,2]. It appears in many situations in hydrodynamics, plasma physics, complex dynamical systems, mathematical biology, nonlinear optics, engineering fields etc. The list is getting longer day by day.

~~~ Integrability provides significant information for physical phenomena described by differential or difference equations. Different analytic and algebraic methods have been developed in this direction [3]. Painlev\'e analysis is an important tool in studying the singularity structure of a differential equation [4]. Discrete Painlev\'e equations have also been constructed following the lineage of continuous ones [5]. Hirota's bilinear method is very much useful for obtaining the multi soliton solution as well as the integrability of nonlinear evolution equations [6]. [7] contains detailed discussions and a summary of other methods.

~~~Exact solutions to nonlinear differential equations may be required not only for their quantitative study but also for validation for certain numerical methods. In the last few decades various methods are proposed to solve nonlinear differential equations analytically. Among them homogeneous balance method [8,9], Tanh method [10], Lie symmetry method [11], B\"acklund transformation method [12], $G'/G$ expansion method [13], Kudryashov method or the $Q-$ function method [14] etc are well known. Homogeneous balance(HB) method are widely used to obtain exact solutions and auto B\"acklund transformations of nonlinear PDEs. Wang et al. [8] used the HB method and obtain new exact solutions of some well known nonlinear PDEs. Fan et al. [15] showed a close relationship among HB method, WTC method and Clarkson-Kruskal (CK) method [16]. Tanh method is used to obtain exact traveling wave solutions to nonlinear PDEs. Malfliet and Hereman developed a systematic version of this method in [10] and by applying this they solved some popular nonlinear PDEs. Lie symmetry method for differential equations was introduced by Sophus Lie. Though it is tedious and highly algorithmic, most of the well known techniques of solving differential equations can be deduced as special cases of the symmetry methods.  

   ~~~In this work our main objective is to find exact solutions of a Boussinesq type equation of  longitudinal wave in a magneto-electro-elastic(MEE) circular rod which was derived by Xue et al.[17]. In recent times, nonlinear dynamics of condensed matter physics, in particular, elastic solids, is an active area of research [18]. MEE structures are used in various engineering fields(like sensors, actuators etc.). Wave propagation in MEE media has got a special value in recent past, many researchers are working on it. Baskonus et al. [19] derived some exact solutions of this equation by modified exponential expansion function method. Ma et al. obtained exact traveling wave solution by modified $G'/G$ expansion method in [20]. The governing equation [17, 19, 20] for longitudinal wave in MEE circular rod is:
\begin{eqnarray}
u_{tt}-v_0^2u_{xx}-(\frac{v_0^2}{2}u^2+mu_{tt})_{xx}=0 \label{1}
\end{eqnarray}
where $v_0$ is the linear longitudinal wave velocity and $m$ is the dispersion parameter. These parameters are dependent on the material and geometry of the rod.

Here we have considered a homogeneous circular rod of infinite length, which is made of composite BaTiO$_3$-CoFe$_2$O$_4$. We consider different material combinations by changing the volume fractions of BaTiO$_3$ as 0\%(PM), 50\%(MEE), 100\% (PE). When the volume of BaTiO$_3$ is 0\% in the rod then it is piezomagnetic(PM), when it is 100\% then it is piezoelectric(PE) and in the case of 50\% BaTiO$_3$ it is magneto-electro-elastic(MEE) [17]. We also considered two purely elastic materials- transversely isotropic elastic material(TI) taking from 50\%(MEE) only the elastic coefficients and effective elastic isotropy(EI) obtained from the TI by making it isotropic. Different values of the linear longitudinal wave velocity $v_0$ and the dispersion parameter $m$ were calculated by Xue et al. in [17]. We listed these values of $v_0, m$ in table 1.


~~~~~~~~ The rest of this paper is arranged as follows. In section 2, Painlev\'e test on $\eqref{1}$ is given. In section 3 and 4, exact solutions  of $\eqref{1}$ are obtained by homogeneous balance and Tanh methods respectively. Section 5 gives a static solution. Section 6 contains Lie symmetry analysis and similarity reduction of \eqref{1}. Group velocity and phase velocity of the corresponding longitudinal waves are found in section 7. Finally, conclusion of this work is drawn in the section 8. 
\section{Painlev\'e Test:}
Painlev\'e test is a powerful tool to check the integrabilty of a differential equation. It was conjectured by Ablowitz et al. [21] that every exact reduction of an  integrable nonlinear partial differential equation to ODE gives rise to an ODE possessing the Painlev\'e property: having no movable singularity other than pole [4]. In other words, the existence of an ODE, not having the Painlev\'e property, by exact reduction from a nonlinear PDE concludes that this PDE is not integrable. In this direction we use the traveling wave variation: 
\begin{eqnarray}
u(x,t)=\psi (z), z=kx-\omega t.
\label{13}
\end{eqnarray}
\eqref{1} becomes
\begin{eqnarray}
m\omega^2 k^2\frac{d^4\psi}{dz^4}+(v_0^2k^2-\omega^2)\frac{d^2\psi}{dz^2}+v_0^2k^2\psi\frac{d^2\psi}{dz^2}+v_0^2k^2(\frac{d\psi}{dz})^2=0
\label{14}
\end{eqnarray}
To perform Painlev\'e test on equation \eqref{1} we collect the leading terms from \eqref{14}
\begin{eqnarray}
m\omega^2 \frac{d^4\psi}{dz^4}+v_0^2\psi\frac{d^2\psi}{dz^2}+v_0^2(\frac{d\psi}{dz})^2=0\label{21}
\end{eqnarray} 
We put $\psi =\frac{a_0}{z^p}$ [14] in \eqref{21} and obtain
\begin{eqnarray}
p=2,\hspace{4mm}a_0=-\frac{m\omega^2}{2v_0^2}
\end{eqnarray}
To find the Fuch's indices [14] we put
\begin{eqnarray}
\psi =\frac{a_0}{z^{-2}}+a_j z^{j}
\end{eqnarray}
in \eqref{21} and equate the coefficients of $a_j$ in the resulting equation to zero and get
\begin{eqnarray}
2j(j-1)(j-2)(j-3)-j(j-1)+4j-6=0\label{22}
\end{eqnarray} 
Solving \eqref{22} we find the Fuch's indices as: $j_1=2, j_2=3, j_3=\frac{1-\sqrt{3}}{2}, j_4=\frac{1+\sqrt{3}}{2}$. The irrational Fuch's indices $j_3, j_4$ indicate that \eqref{14} and hence \eqref{1} fails the Painlev\'e test. Equation \eqref{1} does not possess the Painlev\'e property.
\section{Exact Solution by Homogeneous Balance Method:}
Now we try to find solution of \eqref{1} by homogeneous balance method in the form
\begin{eqnarray}
u(x,t)=\frac{\partial^r}{\partial x^r}f(\phi (x,t))+u_1(x,t)
\end{eqnarray}
where $f, \phi$ are functions of $x,t$ to be determined and $u_1(x,t)$ is a solution of \eqref{1}. Let us take $r=2$.\\ i.e. 
\begin{eqnarray}
u(x,t)=\frac{\partial^2}{\partial x^2}f(\phi (x,t))+u_1(x,t)=f''.\phi_x^2+f'.\phi_{xx}+u_1
\label{9}
\end{eqnarray}
By putting \eqref{9} in the left hand side of \eqref{1} and rearranging we get
\begin{eqnarray}
& &u_{tt}-v_0^2u_{xx}-(\frac{v_0^2}{2}u^2+mu_{tt})_{xx}\nonumber\\& &=-[v_0^2(f'''^2+f''.f^{iv})\phi_x^6 +mf^{vi}\phi_x^4\phi_t^2]-[v_0^2(12f''f'''+f'f^{iv})\phi_x^4 \phi_{xx}+m(8f^{v}\phi_x^3 \phi_{xt} \phi_t \nonumber\\& &+f^{v}\phi_x^4 \phi_{tt}+6f^{v}\phi_x^2 \phi_{xx}\phi_t^2)]+[f^{iv}\phi_x^2 \phi_t^2-v_0^2f^{iv}\phi_x^4-v_0^2(12f''^2\phi_x^2\phi_{xx}^2+2f'f'''\phi_x^3\phi_{xxx}+4f''^2\phi_x^3 \phi_{xxx}\nonumber\\& &+6f'f'''\phi_x^2\phi_{xx}^2+u_1f^{iv}\phi_x^4)-m(12f^{iv}\phi_x^2\phi_{xt}^2+4f^{iv}\phi_x^3\phi_{xtt}+24f^{iv}\phi_x\phi_{xx}\phi_{xt}\phi_t +12f^{iv}\phi_x^2\phi_{xxt}\phi_t\nonumber\\& & +6f^{iv}\phi_x^2\phi_{xx}\phi{tt}+3f^{iv}\phi_{xx}^2\phi_{t}^2+4f^{iv}\phi_x \phi_{xxx}\phi_t^2)]+[4f'''\phi_x \phi_{xt} \phi_t +f'''\phi_x^2 \phi_{tt}+f'''\phi_{xx} \phi_t^2\nonumber\\& & -6v_0^2f'''\phi_x^2 \phi_{xx}-v_0^2(2u_{1x}f'''\phi_x^3 +10f'f''\phi_x \phi_{xx} \phi_{xxx}+f'f''\phi_x^2 \phi_{xxxx}+3f'f'' \phi_{xx}^3+6u_1f''' \phi_x^2 \phi_{xx})\nonumber\\& &-m(12f'''\phi_{xx} \phi_{xt}^2+24f'''\phi_x \phi_{xxt} \phi_{xt}+12f'''\phi_x \phi_{xx} \phi_{xtt}+6f'''\phi_x^2 \phi_{xxtt}+12f'''\phi_{xx} \phi_{xxt} \phi_t \nonumber\\& &+3f'''\phi_{xx}^2 \phi_{tt}+8f'''\phi_{xxx} \phi_{xt} \phi_t +8f'''\phi_x \phi_{xxxt} \phi_t +4f'''\phi_x \phi_{xxx} \phi_{tt} +f'''\phi_{xxxx} \phi_t^2)]+[2f''\phi_{xt}^2\nonumber\\& &+2f''\phi_x \phi_{xtt}+2f''\phi_t \phi_{xxt}+f''\phi_{xx} \phi_{tt}-v_0^2(3f'' \phi_{xx}^2+4f''\phi_x \phi_{xxx})-v_0^2(f'^2\phi_{xxx}^2+6u_{1x}f''\phi_x \phi_{xx}\nonumber\\& &+u_{1xx}f''\phi_x^2+f'^2 \phi_{xx} \phi_{xxxx}+3u_1f'' \phi_{xx}^2+4u_1f''\phi_x \phi_{xxx})-m(6f\phi_{xxt}^2+6f''\phi_{xx} \phi_{xxtt}+8f''\phi_{xxxt} \phi_{xt}\nonumber\\& &+4f''\phi_{xxx} \phi_{xtt}+4f''\phi_x \phi_{xxxtt}+2f''\phi_t \phi_{xxxxt}+f''\phi_{xxxx} \phi_{tt})]+[f'\phi_{xxtt}-v_0^2f'\phi_{xxxx}-v_0^2(2u_{1x}f'\phi_{xxx}\nonumber\\& &+u_{1xx}f'\phi_{xx}+u_1f'\phi_{xxxx})-mf'\phi_{xxxxtt}]+[u_{1tt}-v_0^2u_{1xx}-(\frac{v_0^2}{2}u_1^2+mu_{1tt})_{xx}]\nonumber\\
\label{10}
\end{eqnarray}
Now we equate each third bracketed term in \eqref{10} to zero, to make sure that \eqref{9} gives a solution of \eqref{1}.\\
From the first third bracketed term we have
\begin{eqnarray}
v_0^2(f'''^2+f''.f^{iv})\phi_x^6 +mf^{vi}\phi_x^4\phi_t^2=0
\end{eqnarray}
We assume that 
\begin{eqnarray}
& &f=\sigma \log{\phi},\nonumber\\& &\phi(x,t)=1+\exp(kx-\omega t+\theta_0).
\label{11}
\end{eqnarray}
where $\theta_0$ is arbitrary constant and $k, \omega, \sigma$ are constants to be determined. When we equate each third bracketed term in \eqref{10} to zero, we use \eqref{11} and the results
\begin{eqnarray}
f''f'''=-\frac{\sigma}{12}f^v,\hspace{2mm}f''^2=-\frac{\sigma}{6}f^{iv},\hspace{2mm}f'f'''=-\frac{\sigma}{3}f^{iv},\hspace{2mm}f'f''=-\frac{\sigma}{2}f''',\hspace{2mm}f'^2=-\sigma f''\hspace{4mm} etc. etc.
\end{eqnarray} 
Equating each third bracketed term in \eqref{10} to zero we get a system of algebraic equations. Solving this system we find $u_1$ as a constant and 
\begin{eqnarray}
& &\sigma=\frac{12m\omega^2}{v_0^2k^2}\nonumber\\& &
\omega = \sqrt{\frac{1+u_1}{1-mk^2}}v_0k 
\label{12}
\end{eqnarray} 
Using \eqref{11} and \eqref{12} in \eqref{9} we obtain a solution of \eqref{1}
\begin{eqnarray}
u(x,t)=u_1+\frac{6mk^2(1+u_1)}{(1-mk^2)[1+\cosh(kx-\sqrt{\frac{1+u_1}{1-mk^2}}v_0kt+\theta_0)]}\label{25}
\end{eqnarray}
where $\theta_0, k, u_1$ are constants(can be taken arbitrary).
\begin{figure}[H]
\includegraphics[width=8cm]{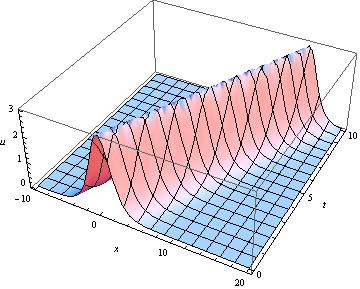} 
\caption{Plot of $u(x,t)$ given in \eqref{25} with $v_0 =k=1, m=\frac{1}{2}, u_1 =\theta_0 =0$}
\end{figure}
\section{Exact Solutions by Tanh Method:}
We are now going to find solution of \eqref{1} by Tanh method.
We are looking for the solution of \eqref{14} with the boundary condition 
\begin{eqnarray}
\psi(z)\longrightarrow 0,\hspace{5mm} \frac{d^n\psi}{dz^n}\longrightarrow 0\hspace{2mm}(n=1, 2, 3,...)\hspace{5mm} for\hspace{5mm} z\longrightarrow \pm \infty.
\label{17}
\end{eqnarray}  
With the assumption that traveling wave solutions are expressible in terms of $\tanh{z}$, we introduce $Y=\tanh{z}$ as a new dependent variable. Let
\begin{eqnarray}
u(x,t)=\psi(z)=S(Y)=\sum_{n=0}^{N}a_nY^n,\hspace{5mm}Y=\tanh{z}=\tanh{(kx-\omega t)}
\label{15}
\end{eqnarray}
Equation \eqref{14} becomes   
\begin{eqnarray}
& &m\omega^2 k^2(1-Y^2)^4\frac{d^4S}{dY^4}-12m\omega^2 k^2Y(1-Y^2)^3\frac{d^3S}{dY^3}+\{(v_0^2k^2-\omega^2)(1-Y^2)^2+v_0^2k^2S(1-Y^2)^2\nonumber\\& &-2m\omega^2 k^2(1-Y^2)^2(4-18Y^2)\}\frac{d^2S}{dY^2}+v_0^2k^2(1-Y^2)^2\left(\frac{dS}{dY}\right)^2+2\{2m\omega^2 k^2(4-6Y^2)\nonumber\\& &-(v_0^2k^2-\omega^2)-v_0^2k^2S\}Y(1-Y^2)\frac{dS}{dY}=0
\label{16}
\end{eqnarray}
Using \eqref{15} in \eqref{16} and balancing the highest degree in $Y$ we get $N=2$. The boundary condition \eqref{17} implies
\begin{eqnarray}
S(Y)\longrightarrow 0\hspace{4mm} for\hspace{4mm}Y\longrightarrow \pm 1
\end{eqnarray}
We only consider the limit $Y\longrightarrow 1$. Then the possible form of the solution is
\begin{eqnarray}
S(Y)=b_0(1-Y)(1+b_1Y)
\label{18}
\end{eqnarray}
where $b_0, b_1$ are constants to be determined (remember that $N=2$ and $S(Y)\longrightarrow 0$ as $Y\longrightarrow 1$).\\
We put \eqref{18} in \eqref{16}, then equate the coefficients of different powers $Y$ to zero and obtain
\begin{eqnarray}
b_0=\frac{12m\omega^2}{v_0^2},\hspace{4mm}b_1=1,\hspace{4mm}k=\pm \frac{\omega}{\sqrt{4m\omega^2 +v_0^2}}
\label{19}
\end{eqnarray}
Using \eqref{19} in \eqref{18} we obtain solutions of \eqref{1} as:
\begin{eqnarray}
u(x,t)=\frac{12m\omega^2}{v_0^2} Sech^2{\left(\pm \frac{\omega}{\sqrt{4m\omega^2 +v_0^2}}x-\omega t\right)}\label{26}
\end{eqnarray} 
where $\omega$ is any constant.
\begin{figure}[H]
\includegraphics[width=8cm]{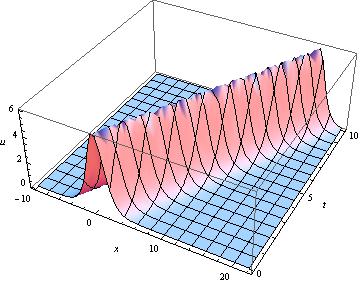} 
\caption{Plot of $u(x,t)$ given in \eqref{26} with + sign and $v_0 =\omega=1, m=\frac{1}{2}$}
\includegraphics[width=8cm]{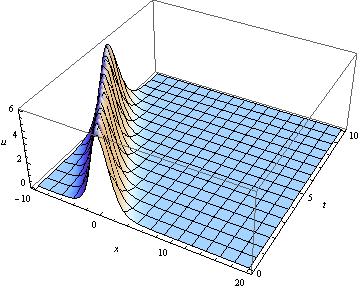} 
\caption{Plot of $u(x,t)$ given in \eqref{26} with - sign and $v_0 =\omega=1, m=\frac{1}{2}$}

\end{figure}
\section{Static Solutions:}
Now we are going to find a solution of \eqref{1} in the form
\begin{eqnarray}
u(x,t)=cf(x),\label{20}
\end{eqnarray} 
where $c$ is a constant and $f$ is a function to be determined.\\
We put \eqref{20} in \eqref{1} and integrating get
\begin{eqnarray}
f(x)=\frac{-1\pm \sqrt{1+2(Ax+B)c}}{c}\nonumber
\end{eqnarray} 
Thus we obtain
\begin{eqnarray}
u(x,t)=-1\pm \sqrt{1+2(Ax+B)c},\label{27}
\end{eqnarray}
a static solution of \eqref{1}, where $A, B, c$ are arbitrary constants. 
\begin{figure}[H]
\includegraphics[width=8cm]{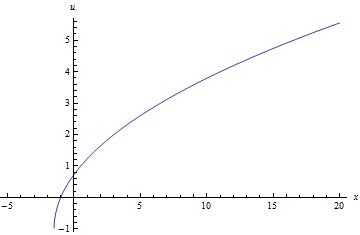} 
\caption{Plot of $u(x,t)$ given in \eqref{27} with + sign and $v_0 =A=B=c=1$}
\includegraphics[width=8cm]{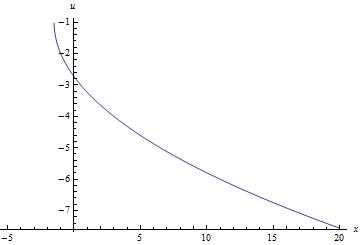} 
\caption{Plot of $u(x,t)$ given in \eqref{27} with - sign and $v_0 =A=B=c=1$}

\end{figure}

\section{Lie Symmetry Analysis and Similarity Reduction:}

We want to solve \eqref{1} by using the Lie symmetry method. We introduce the symmetry transformations as:
$(x,t,u)\rightarrow(\hat{x},\hat{t},\hat{u})$ where
\begin{eqnarray}
\hat{x}=\hat{x}(x,t,u(x,t);\epsilon),
\hat{y}=\hat{y}(x,t,u(x,t);\epsilon),
\hat{z}=\hat{z}(x,t,u(x,t);\epsilon).
\end{eqnarray} 
Tangent vectors at $(\hat{x},\hat{t},\hat{u})$ is $(\xi,\tau,\eta)$ where 
\begin{eqnarray}
\frac{d\hat{x}}{d\epsilon}=\xi(\hat{x},\hat{t},\hat{u}),\frac{d\hat{t}}{d\epsilon}=\tau(\hat{x},\hat{t},\hat{u}),\frac{d\hat{u}}{d\epsilon}=\eta(\hat{x},\hat{t},\hat{u})
\end{eqnarray}
subject to the initial condition 
\begin{eqnarray}
(\hat{x},\hat{t},\hat{u})|_{\epsilon=0}=(x,t,u)
\end{eqnarray}
We want point symmetries of the form
\begin{eqnarray}
\hat{x}=x+\epsilon \xi(x,t,u)+o(\epsilon^2)\\
\hat{t}=t+\epsilon \tau(x,t,u)+o(\epsilon^2)\\
\hat{u}=u+\epsilon \eta(x,t,u)+o(\epsilon^2)
\end{eqnarray}
The prolongation of the above point transformation to derivatives are given below:
\begin{eqnarray}
& &\hat{u}_{\hat{x}}=u_x+\epsilon\eta^x(x,t,u,u_x,u_t)+o(\epsilon^2)\nonumber\\& &
\hat{u}_{\hat{t}}=u_t+\epsilon\eta^t(x,t,u,u_x,u_t)+o(\epsilon^2)\nonumber\\& &
\hat{u}_{\hat{x}\hat{x}}=u_{xx}+\epsilon\eta^{xx}(x,t,u,u_x,u_t,...)+o(\epsilon^2)\nonumber\\& &
\hat{u}_{\hat{t}\hat{t}}=u_{tt}+\epsilon\eta^{tt}(x,t,u,u_x,u_t,...)+o(\epsilon^2)\nonumber\\& &
\hat{u}_{\hat{t}\hat{t}\hat{x}\hat{x}}=u_{ttxx}+\epsilon\eta^{ttxx}(x,t,u,u_x,u_t,...)+o(\epsilon^2)\nonumber\\
\label{2}
\end{eqnarray}
where
\begin{eqnarray}
& &\eta^x=D_x\eta-u_xD_x\xi-u_tD_x\tau\nonumber\\& &
\eta^t=D_t\eta-u_xD_t\xi-u_tD_t\tau\nonumber\\& &
\eta^{xx}=D_x\eta^x-u_{xx}D_x\xi-u_{xt}D_x\tau\nonumber\\& &
\eta^{tt}=D_t\eta^t-u_{tx}D_t\xi-u_{tt}D_t\tau\nonumber\\& &
\eta^{ttx}=D_x\eta^{tt}-u_{ttx}D_x\xi-u_{ttt}D_x\tau\nonumber\\& &
\eta^{ttxx}=D_x\eta^{ttx}-u_{ttxx}D_x\xi-u_{tttx}D_x\tau\nonumber\\& &
\label{4}
\end{eqnarray}
and $D_x=\partial_x+u_x\partial_u+u_{xx}\partial_{u_x}+u_{tx}\partial_{u_t}+u_{xxt}\partial_{u_{xt}}+...$\\
We rewrite \eqref{1} as
\begin{eqnarray}
u_{tt}-v_0^2u_{xx}-v_0^2u_{x}^2-v_0^2uu_{xx}-mu_{ttxx}=0
\end{eqnarray}
Thus, the symmetry condition for \eqref{1} is
\begin{eqnarray}
& &\hat{u}_{\hat{t}\hat{t}}-v_0^2\hat{u}_{\hat{x}\hat{x}}-v_0^2\hat{u}_{\hat{x}}^2-v_0^2\hat{u}\hat{u}_{\hat{x}\hat{x}}-m\hat{u}_{\hat{t}\hat{t}\hat{x}\hat{x}}=0\nonumber\\& &
\Rightarrow u_{tt}+\epsilon\eta^{tt}-v_0^2(u_{xx}+\epsilon\eta^{xx})-v_0^2(u_x+\epsilon\eta^x)^2-v_0^2(u+\epsilon\eta)(u_{xx}+\epsilon\eta^{xx})-m(u_{ttxx}+\epsilon\eta^{ttxx})\nonumber\\& &+o(\epsilon^2)=0   \hspace{10mm}  [by \eqref{2}]
\label{3}
\end{eqnarray}
Equating the coefficients of $\epsilon$ in \eqref{3} to zero we get the linearized symmetry condition:
\begin{eqnarray}
\eta^{tt}-v_0^2\eta^{xx}-2v_0^2u_x\eta^x-v_0^2(u\eta^{xx}+u_{xx}\eta)-m\eta^{ttxx}=0
\label{5}
\end{eqnarray}
We find the explicit values of $\eta^x, \eta^{xx}, \eta^{tt}, \eta^{ttxx}$ from \eqref{4} and then put these values in \eqref{5}. Since $\xi, \tau, \eta$ are independent of the derivatives of $u$ then we equate the coefficients of different derivatives of $u$ from both sides of the resulting equation. Simplifying we get:
\begin{eqnarray}
\xi =c_1, \tau =-c_2\frac{t}{2}+c_3, \eta =c_2(1+u).
\end{eqnarray}
where $c_1, c_2, c_3$ are arbitrary constants. The infinitesimal symmetry generator is given by:
\begin{eqnarray}
X=\xi \frac{\partial}{\partial x}+\tau \frac{\partial}{\partial t}+\eta \frac{\partial}{\partial u}=c_1 \frac{\partial}{\partial x}+c_2\left(-\frac{t}{2}\frac{\partial}{\partial t}+\frac{\partial}{\partial u}+u\frac{\partial}{\partial u}\right)+c_3\frac{\partial}{\partial t}
\end{eqnarray}
The Lie algebra of the point symmetry generators is spanned by:
\begin{equation}
X_1=\frac{\partial}{\partial x}, X_2=-\frac{t}{2}\frac{\partial}{\partial t}+\frac{\partial}{\partial u}+u\frac{\partial}{\partial u}, X_3=\frac{\partial}{\partial t}
\end{equation}  
Now every invariant solution satisfies the invariant surface condition: 
\begin{eqnarray}
\eta-\xi u_x-\tau u_t=0
\end{eqnarray}
which is quasilinear partial differential equation having Lagrange's auxiliary equations: 
\begin{equation}
\frac{dx}{\xi}=\frac{dt}{\tau}=\frac{du}{\eta}
\label{6}
\end{equation}
Solving \eqref{6} with $c_1=1, c_2=1$ and $c_3$ arbitrary we get
\begin{eqnarray}
u=\frac{G(x+2\log(c_3-\frac{t}{2}))}{(c_3-\frac{t}{2})^2}-1
\label{7}
\end{eqnarray}
where $G$ is an arbitrary function of single variable. By putting this value of $u$ in \eqref{1} we get the following ordinary differential equation:
\begin{eqnarray}
& &-3G(r)+5G'(r)+2v_0^2{G'(r)}^2-2G''(r)+3mG''(r)+2v_0^2G(r)G''(r)-5mG'''(r)\nonumber\\& &+2mG^{iv}(r)=0
\label{8}
\end{eqnarray}
where $r=x+2\log(c_3-\frac{t}{2})$.\\
Therefore, \eqref{7} gives a solution of \eqref{1}, provided $G$ satisfies \eqref{8}. Equation \eqref{8} represents a similarity reduction of \eqref{1}. Here we present some numerical solutions of \eqref{8} for some specific values of the parameters $v_0, m$ given in the table 1 :\\
\begin{figure}[H]
\includegraphics[width=8cm]{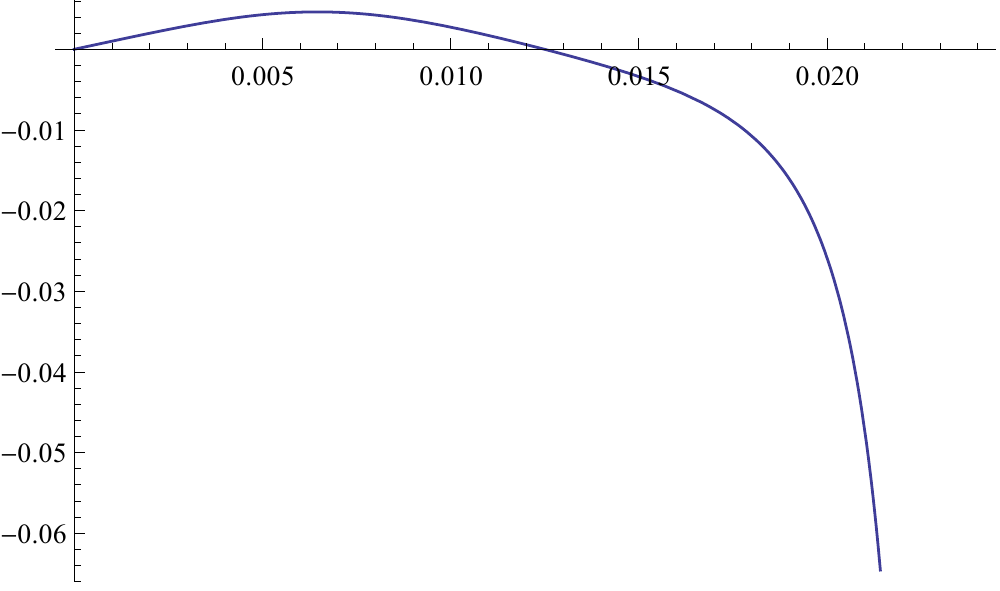} 
\caption{Plot of $G(r)$ given in \eqref{8} with $v_0=5213.1, m=1.7350\times 10^{-4}$}
\includegraphics[width=8cm]{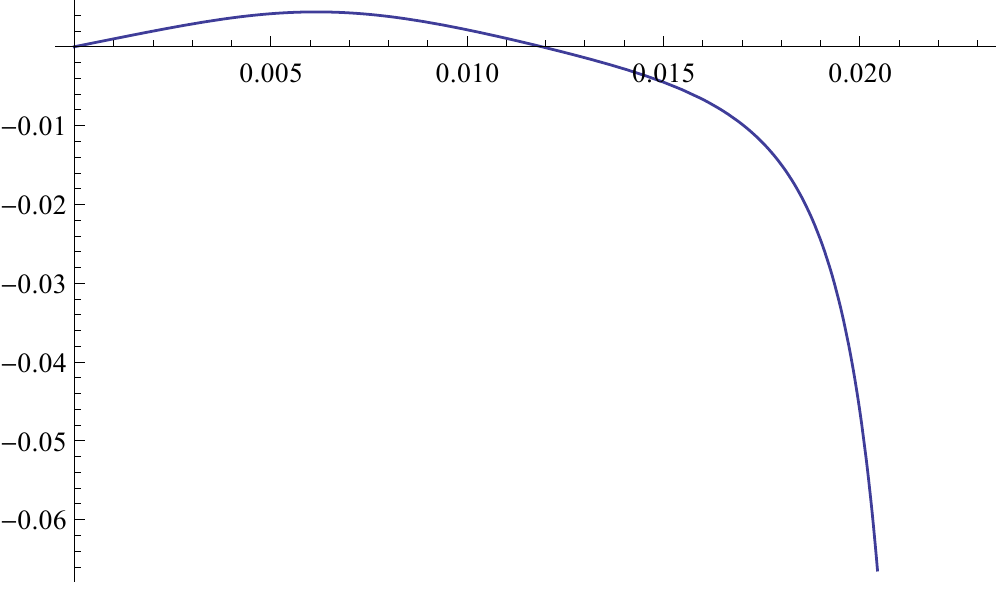} 
\caption{Plot of $G(r)$ given in \eqref{8} with $v_0=5144.6, m=1.4890\times 10^{-4}$}
\includegraphics[width=8cm]{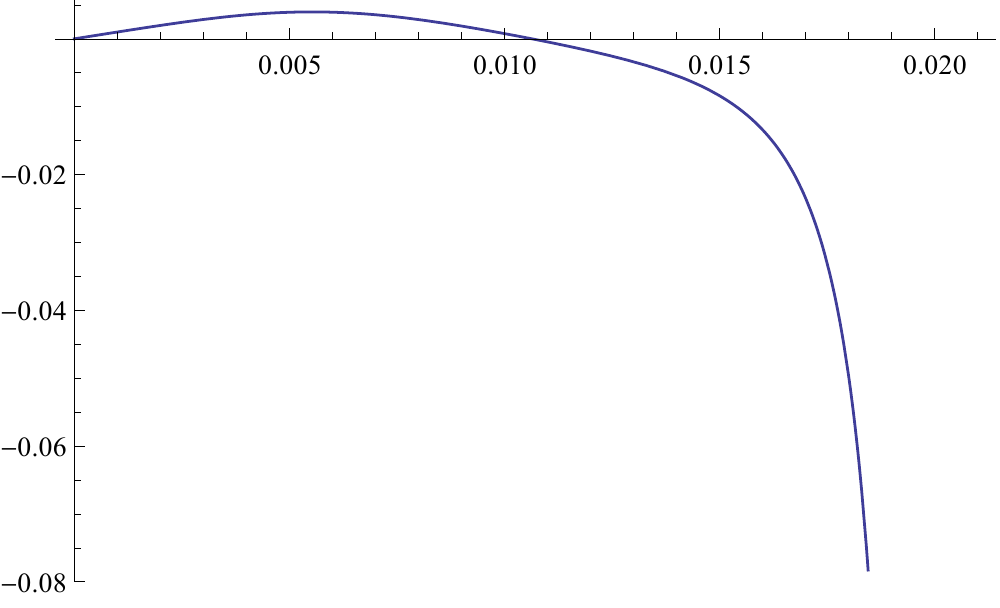} 
\caption{Plot of $G(r)$ given in \eqref{8} with $v_0=5049.8, m=1.0560\times 10^{-4}$}
\end{figure}
\begin{figure}[H]
\includegraphics[width=8cm]{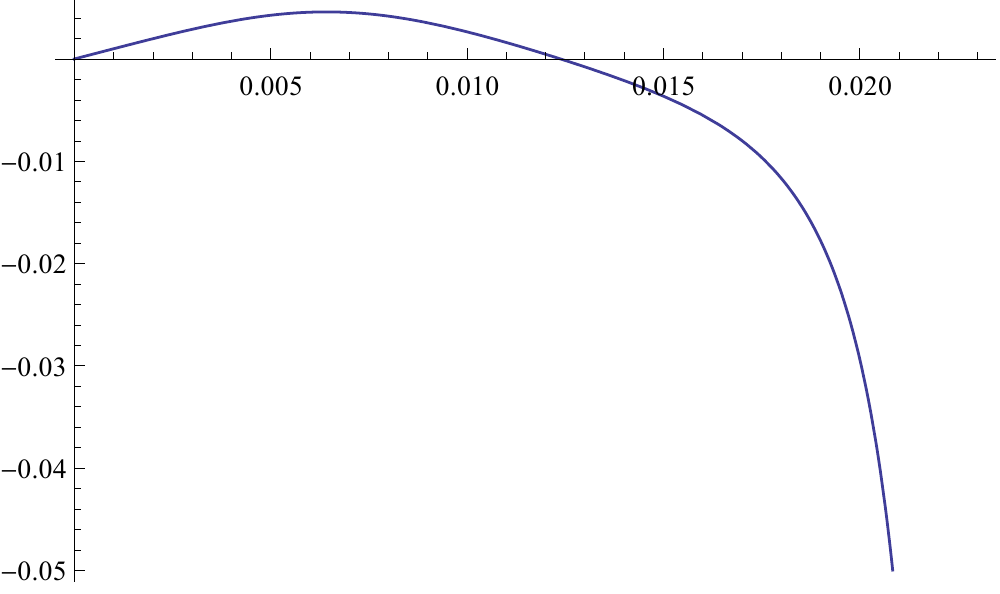} 
\caption{Plot of $G(r)$ given in \eqref{8} with $v_0=4800.3, m=1.5700\times 10^{-4}$}
\includegraphics[width=8cm]{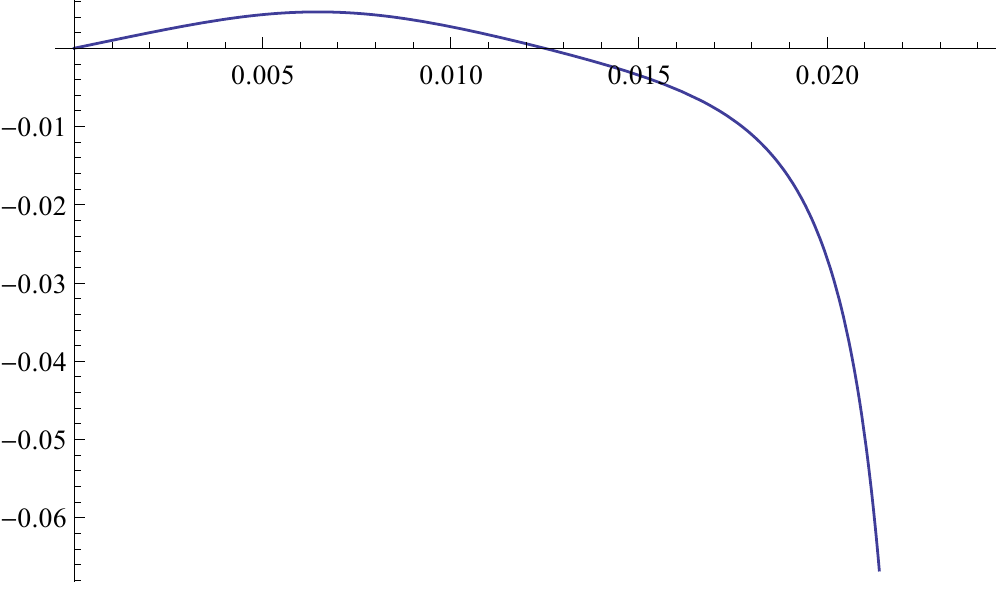} 
\caption{Plot of $G(r)$ given in \eqref{8} with $v_0=4839.8, m=1.6200\times 10^{-4}$}
\end{figure}
\section{Group Velocity, Phase Velocity:}
To obtain the group velocity, phase velocity of \eqref{1} we use the perturbation technique. We put $u(x,t)=u_0+u_1(x,t)$ in \eqref{1}, where $u_0$ is a constant equilibrium state and $u_1(x,t)$ is the perturbation. \eqref{1} becomes
\begin{eqnarray}
u_{1tt}-v_0^2u_{1xx}-(v_0^2u_0u_{1xx}+mu_{1ttxx})=0\label{23}
\end{eqnarray}
To obtain the phase velocity, group velocity we take $u_1(x,t)=\exp{\{i(kx-\omega t)\}}$. Then clearly $\frac{\partial}{\partial t}\equiv -i\omega$ and $\frac{\partial}{\partial x}\equiv ik$. Using these results in  \eqref{23} we get 
\begin{eqnarray}
\omega^2 =\left(\frac{1+u_0}{1+mk^2}\right) v_0^2k^2\label{24}
\end{eqnarray}
Thus we obtain the phase velocity as
\begin{eqnarray}
v_p =\frac{\omega}{k}=\pm \sqrt{\frac{1+u_0}{1+mk^2}}v_0
\end{eqnarray}
which depends on the wave number $k$, the wave is dispersive.\\ From \eqref{24} we obtain the group velocity
\begin{eqnarray}
v_g =\pm \frac{1}{1+mk^2} \sqrt{\frac{1+u_0}{1+mk^2}}v_0=\frac{1}{1+mk^2} v_p
\end{eqnarray}
\begin{figure}[H]
\includegraphics[width=8cm]{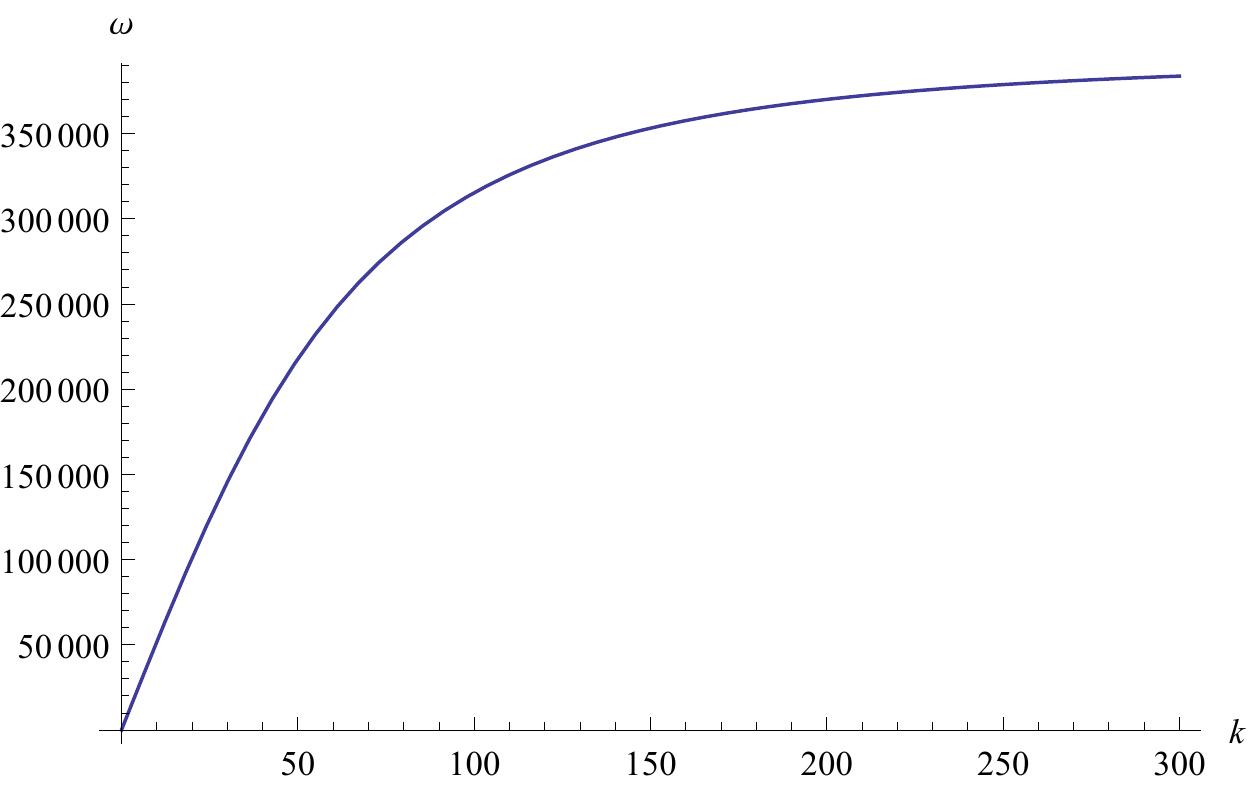} 
\caption{Relation between frequency $\omega$ and wave no. $k$ with $v_0=5213.1,m=1.735\times 10^{-4},u_0=0$}
\end{figure}
\begin{figure}[H]
\includegraphics[width=8cm]{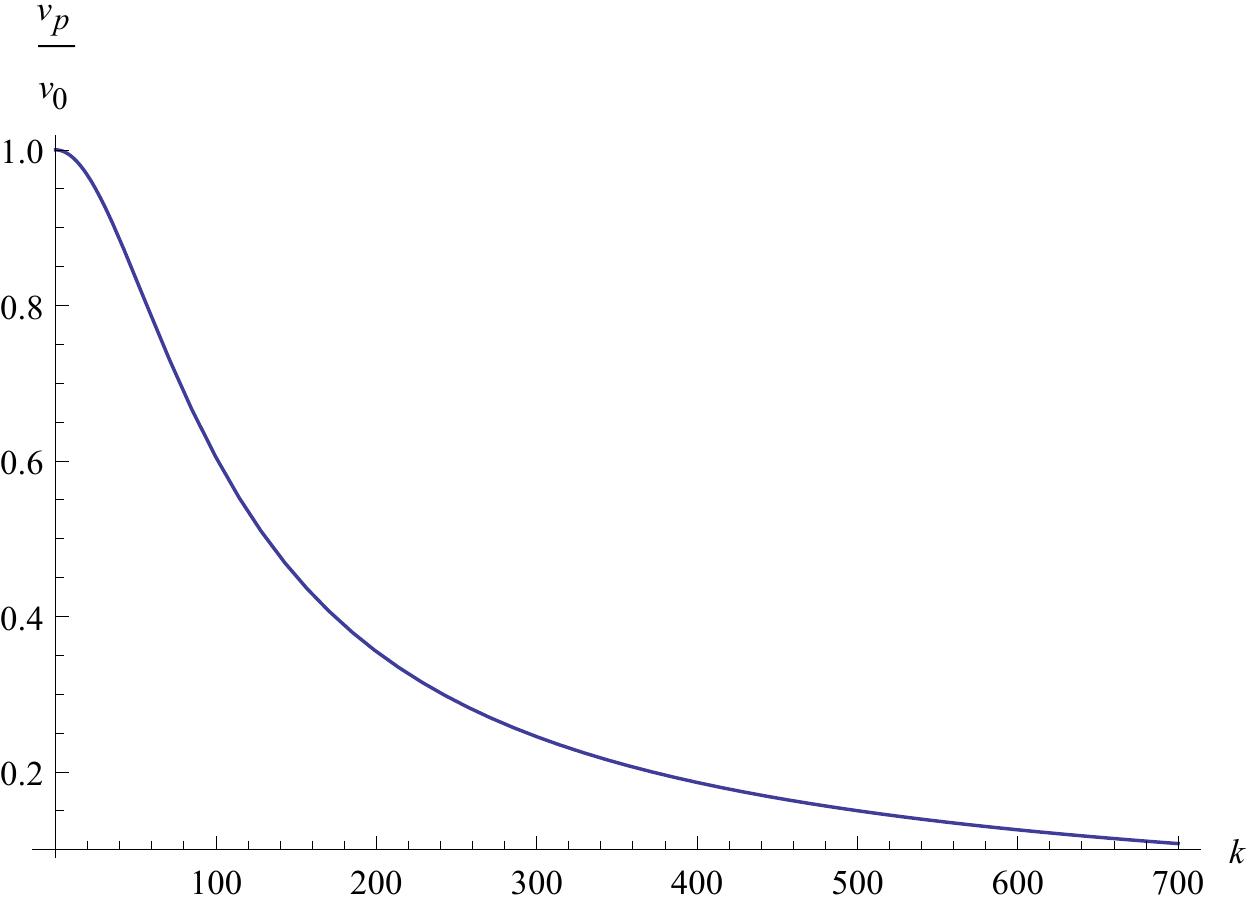} 
\caption{Relation between normalized phase velocity $\frac{v_p}{v_0}$ and wave no. $k$ with $m=1.735\times 10^{-4},u_0=0$}
\includegraphics[width=8cm]{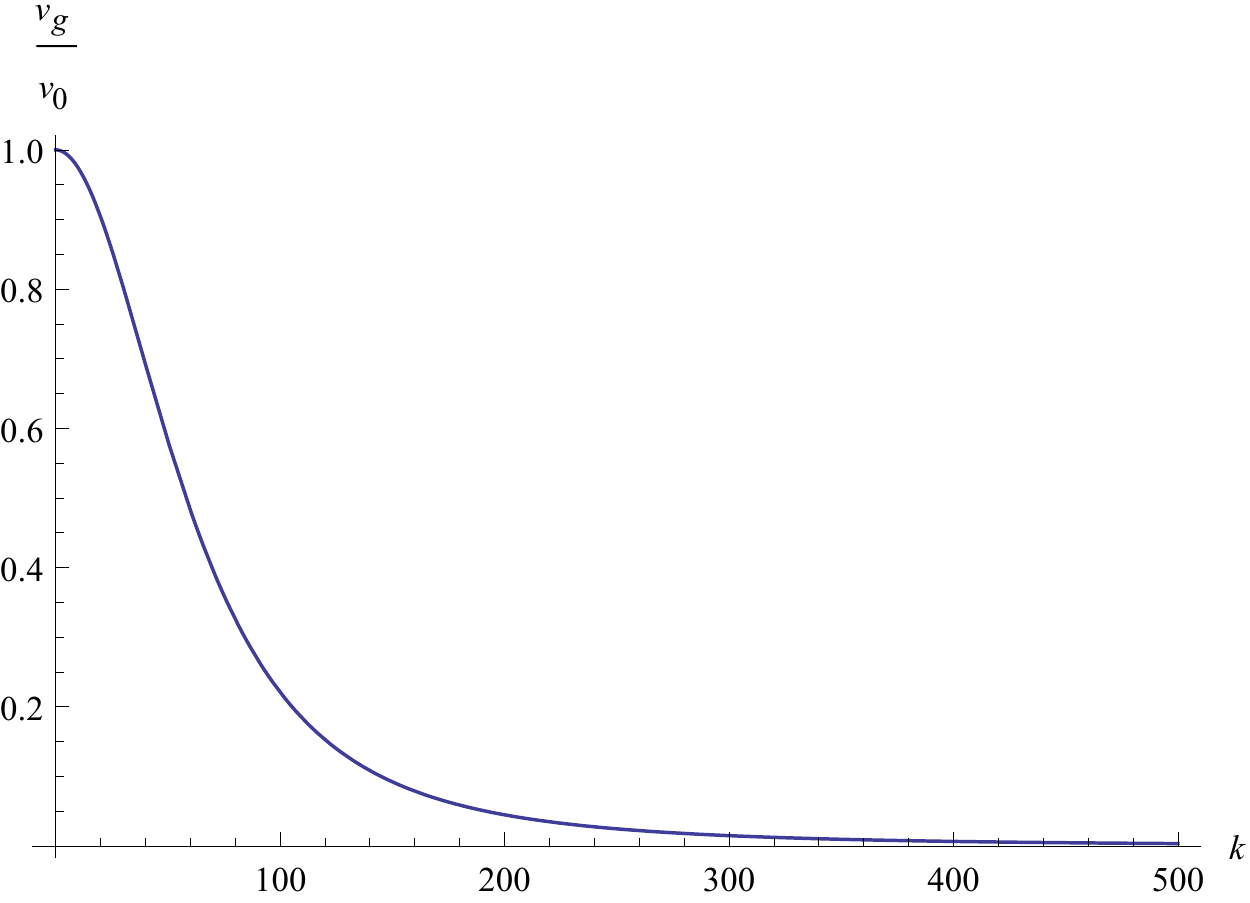} 
\caption{Relation between normalized group velocity $\frac{v_g}{v_0}$ and wave no. $k$ with $m=1.735\times 10^{-4},u_0=0$}
\end{figure}
Here we give some numerical values of $v_p, v_g$ for some particular values of $v_0, m$ (given in table 1)
\begin{table}[h!]
\begin{center}
\caption{Numerical Values of $v_p, v_g$ for some particular $v_o, m$ and $k=5,u_0=0$}
\begin{tabular}{|c|c|c|c|c|}
\hline
 \textbf{Volume fraction of BaTiO$_3$}& \textbf{$v_0$}&\textbf{$m$} &\textbf{$v_p$}&\textbf{$v_g$}\\
 \hline
 0\%(PM)& 5213.1&0.0001735&5201.83&5179.37\\
 50\%(MEE)& 5144.6&0.0001489&5135&5116.01\\
 100\%(PE)&5049.8&0.0001056&5043.15&5029.87\\
 Transverse isotropy (TI)&4800.3&0.000157&4790.91&4772.18\\
 Elastic isotropy (EI)&4839.8&0.000162&4830.03&4810.55\\
 \hline

\end{tabular}
\end{center}
\end{table}
\section{Conclusion:}
Here a Boussinesq type equation for longitudinal waves through MEE circular rod has been investigated. The equation does not satisfy the Painlev\'e criterion for integrability. Lie point symmetry generators are obtained explicitly, one similarity reduction is found and solved numerically. The obtained solitary wave solutions are plotted. A solution very similar to \eqref{26} was found in [17] by Jacobi elliptic function method and also found in [20] by $G'/G$ expansion method. The solution \eqref{25} is a new solution, as per our knowledge it was not found earlier. The exactness of the solutions are checked by Mathematica software. The normalized group and phase velocity of the original dispersion equation are plotted with respect to wave number $k$. The numerical values of phase velocity and group velocity are calculated for different materials in table 1. We have observed that in the coupled class (PM, MEE, PE) both the group and phase velocities attain their highest value for PM and lowest value in PE. i.e. phase and group velocities decrease with increment of the volume fraction of BaTiO$_3$ in the composite material of the rod. These velocities are higher in coupled class than in the purely elastic class (EI and TI). 
\section{Acknowledgement:}
Arindam Ghosh is grateful to MHRD India for their financial support.

\textbf{Conflict of Interest:} The authors declare that they have no conflict of interest.\\

\textbf{Data Availability Statement:} The data that supports the findings of this study are available within the article.

\section{References:}
$[1]$ Jordan D.W., Smith P.: Nonlinear ordinary differential equations. Oxford University Press (2007).\\
$[2]$ Strogatz S.H.: Nonlinear Dynamics and Chaos.  Perseus Books (1994).\\
$[3]$ Wang Y., Chen Y.: Journal of mathematical physics, 53, 123504(2012).\\
$[4]$Conte. R., Musette. M: The Painlev\'e handbook. Springer, 2008.\\
$[5]$ Brezin E., Kazakov V.A.: Phys. Lett. 236B,(1990)144 \\
$[6]$ Hirota R., phys. rev. lett.27(1971),1192\\
$[7]$ Drazin P.G., Johnson R.S.: Soliton, Chembridge University Press, 1989.\\ 
$[8]$ Wang. M., Zhou. Y., Li.Z.: Application of a homogeneous balance method to exact solutions of nonlinear equations in mathematical physics. Physics Letters A, 216(1-5), 67-75(1996).\\
$[9]$ Maitra. S., Ghosh. A., Roy Chowdhury. A.: Exact solutions and symmetry analysis of a new equation invariant under scaling of dependent variable. Physica Scripta, 94(085212), 2019.\\
$[10]$ Malfliet. W., Hereman. W.: The tanh method: I. Exact solutions of nonlinear evolution and wave equations. Physica Scripta, 54(563-568), 1996.\\ 
$[11]$ Hydon. P.E.: Symmetry methods for differential equations. Cambridge University Press, 2000.\\
$[12]$ Rogers. C., Shadwick. W.F.: B\"acklund Transformations and their applications, Academic Press, 1982.\\
$[13]$ Manafian, J., Lakestani, M., A new analytical approach to solve some fractional-order partial differential equation. Indian J Phys 91,243-258(2017).\\
$[14]$ Kudryashov N.A.,Zakharchenko A.S.: Analytical properties and exact solutions of the  Lotka-Volterra competition system. Applied mathematics and computation, 254(219-228), 2015. \\
$[15]$Fan E.: Two new applications of the homogeneous balance method. Physics Letters A, 265, 353-357(2000).\\ 
$[16]$Clarkson P. A., Kruskal M. D.: New similarity reductions of the Boussinesq equation. Journal of mathematical physics, 30, 2201(1989).\\
$[17]$ Xue C.X., Pan E., Zhang S.Y.: Solitary waves in a magneto-electro-elastic circular rod. Smart Mater. Struct. 20, 105010(2011).\\
$[18]$Samsonov A. M.: Strain Solitons in Solids and How to Construct Them. Chapman and Hall/CRC, 2001.\\ 
$[19]$ Baskonus H.M., Bulut H., Atangana A.: On the complex and hyperbolic structures of the longitudinal wave equation in a magneto-electro-elastic circular rod. Smart Mater. Struct. 25, 035022(2016).\\
$[20]$ Ma X., Pan Y., Chang L.: Explicit traveling wave solutions in a magneto-electro-elastic circular rod. Int. J. Comput. Sci. Issues 10(1), 62-68(2013).\\
$[21]$M.J. Ablowitz, A. Ramani, H. Segur: A connection between nonlinear evolution equations and ordinary differential equations of P-type I. J.Math.Phys., 21(1980),715.
\end{document}